\documentclass{mn2e}
\usepackage{graphicx,times}

\def \apj{ApJ}

\def \mnras{MNRAS}

\title[Comparing dynamical and spectroscopic IMF results]{Variations in the initial mass function in early-type galaxies:\\
A critical comparison between dynamical and spectroscopic results}

\author[Russell J. Smith]{Russell J. Smith\\Department of Physics, University of Durham, South Road, Durham \ DH1 3LE}

\pagerange{\pageref{firstpage}$-$\pageref{lastpage}} \pubyear{2014}

\begin{document}

\label{firstpage}

\voffset=-1.5cm

\maketitle

\begin{abstract}
I present a comparison between published dynamical (ATLAS3D) and spectroscopic (Conroy \& van Dokkum) 
constraints on the stellar initial mass function (IMF) in early-type galaxies, using the 34 galaxies in common between the two works.
Both studies infer an average IMF mass factor  $\alpha$ (the stellar mass relative to a Kroupa-IMF population of similar age and metallicity)
greater than unity, i.e. both methods favour an IMF which is heavier than 
that of the Milky Way, {\it on average} over the sample. 
However, on a galaxy-by-galaxy basis, there is no correlation between $\alpha$ inferred from the two approaches. 
I investigate how the two estimates of $\alpha$ are correlated systematically with the galaxy velocity 
dispersion, $\sigma$, and with the Mg/Fe abundance ratio.
The spectroscopic method, based on the strengths of metal absorption lines, yields a correlation
only with metal abundance ratios: at fixed Mg/Fe, there is no residual correlation with $\sigma$.
The dynamical method, applied to exactly the same galaxy sample, yields the {\it opposite} result:
the IMF variation correlates only with dynamics, with no residual correlation with Mg/Fe
after controlling for $\sigma$.
Hence although both methods indicate a heavy IMF on average in ellipticals,
they lead to incompatible results for the systematic trends, when applied to the same set of galaxies.
The sense of the disagreement could suggest that one (or both) of the methods has not accounted fully for the main confounding 
factors, i.e. element abundance ratios or dark matter contributions. 
Alternatively, the poor agreement might indicate additional variation in the detailed shape of the IMF, beyond what can currently be inferred from the 
spectroscopic features.
\end{abstract}
\begin{keywords}
galaxies: stellar content ---
galaxies: elliptical and lenticular, cD
\end{keywords}

\section{Introduction}

\begin{figure*}
\includegraphics[angle=270,width=175mm]{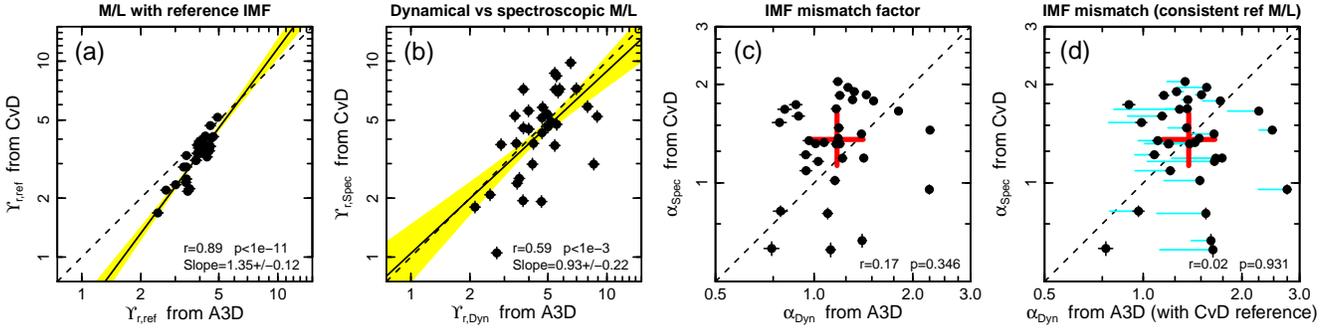}
\vskip -1mm
\caption{Comparison of the mass-to-light ratios, $\Upsilon$, used by CvD and A3D, and the derived mismatch parameters, $\alpha$. 
Panel ``a'' compares the reference mass-to-light ratios, i.e. those derived from stellar population fitting assuming a standard (Kroupa 2001) IMF.
Panel ``b''  compares the derived spectroscopic and dynamical mass-to-light ratios. Panel ``c" compares the ratios $\alpha$\,=\,$\Upsilon/\Upsilon_{\rm ref}$,
which indicate the inferred deviations from the standard IMF. In panel ``d'',  the A3D value of $\alpha$ is modified by using the same $\Upsilon_{\rm ref}$ as
used by CvD, for greater consistency (cyan lines show the effect of this change). The red cross in panels ``c'' and ``d'' shows the median value for the two datasets.
The quoted slopes and correlation coefficients $r$ are for comparisons in  $\log\Upsilon$ or $\log\alpha$; $p$ is the probability of a larger $r$ under the null hypothesis 
of no correlation. In panels ``a'' and ``b'' the solid line and yellow shading show a linear fit with errors.
While the mass-to-light ratios are well correlated between the studies, the ratio $\alpha$ shows essentially no correlation on a galaxy by galaxy basis.}
\label{fig:mlcomp}
\end{figure*}

The stellar initial mass function (IMF) is a key quantity in astrophysics, both intrinsically, as a constraint on the physics of star formation, and also 
for its importance in converting observed galaxy luminosities into physically-meaningful stellar masses, star-formation rates, etc.
In recent years several largely-independent methods have found evidence for a different IMF in early-type galaxies, compared to the Milky Way (MW), 
and for systematic variation among elliptical galaxies as a function of their mass (Treu et al. 2010; 
Auger et al. 2010; van Dokkum \& Conroy 2010; Conroy \& van Dokkum 2012b; Smith, Lucey \& Carter 2012; 
Spiniello et al. 2012; La Barbera et al. 2013; Cappellari et al. 2013). 

One method to constrain the IMF uses measurements of spectral features that are sensitive to surface gravity at fixed stellar temperature, and hence reveal the 
presence of low-mass stars in integrated-light spectra (Conroy \& van Dokkum 2012a).
Dwarf-star-sensitive features are found to increase in strength in higher mass galaxies, beyond what is expected from element
abundance trends, according to spectral synthesis models. This behaviour is interpreted as due to an increasingly bottom-heavy IMF in higher mass galaxies. 
As noted by Conroy \& van Dokkum (2012b), the signature of bottom-heavy IMFs appears more strongly correlated with 
Mg/Fe abundance ratios than with velocity dispersion, suggesting
that IMF could be linked to star-formation intensity, with more low-mass stars being produced in rapid bursts. 

Another technique is to infer the total mass in galaxies from
gravitational tracers, such as stellar dynamics (e.g. Cappellari et al. 2013) or strong lensing 
(e.g. Treu et al. 2010). 
After accounting for the dark matter halo  contribution, this leads to an estimate of stellar mass-to-light ratio $\Upsilon$. Comparing 
this to the ``reference'' mass-to-light  ($\Upsilon_{\rm ref}$) expected from the spectrum of the galaxy assuming a MW-like IMF, 
this yields a mismatch factor $\alpha$\,=\,$\Upsilon/\Upsilon_{\rm ref}$.
Lensing and dynamical studies both find a trend of increasing $\alpha$ with galaxy mass, which can be attributed to an increasing 
contribution of low-mass stars, i.e. a more bottom-heavy IMF.

It should be stressed that these two approaches to constraining the IMF measure fundamentally different quantities: 
gravitational tracers strictly measure mass (which could be dominated by very low-mass dwarfs, or by remnants from massive stars),
while the spectroscopic method is sensitive only to low-mass stars. When mass-to-light ratios are quoted from spectroscopy, as by CvD,
the values depend on a model assumed for the shape of the IMF. Since spectroscopy and dynamics/lensing each measures a differently-weighted
integral over the IMF, comparing the two methods yields a test for the correctness and universality of the assumed IMF model,
as well as a test for the systematic errors inherent to each method.
For example La Barbera et al. (2013) emphasise that although single- and broken-power-law IMFs can fit their spectroscopic data equally well,
the best fitting single-power-law model can be excluded, since it would imply an excessively high $\Upsilon$ for the most massive galaxies. 

At face value, the recent spectroscopic and dynamical/lensing results do appear to agree, at least at a qualitative level: massive early-type galaxies have 
IMFs which, on average, are more bottom-heavy than that of the MW, and there is a trend of increasing deviation from the MW IMF at larger mass. 
This apparent consensus between largely-independent methods has understandably led to increased confidence in these results\footnote{But note that 
some other works have favoured MW-like IMFs even in very massive ellipticals, using lensing (Smith \& Lucey 2013) and dynamics (J. Thomas et al. in preparation, see  
http://www.mpa-garching.mpg.de/halo2013/pdfs/day5/11\_thomas.pdf).}.
In this Letter, I present a critical evaluation of results obtained by Conroy \& van Dokkum (2012b) and by Cappellari et al. (2013) (hereafter CvD and A3D),
where comparisons can be made for exactly the same set of galaxies. I start by directly comparing the spectroscopic and dynamical mass-to-light ratios, 
the reference mass-to-light ratios, and the IMF mismatch factors, on a galaxy-by-galaxy basis (Section~\ref{sec:mlcomp}). Section~\ref{sec:imftrends}  
investigates the systematic correlations of mismatch factor with velocity dispersion and Mg/Fe ratios. In Section~\ref{sec:disc}, I highlight the very different 
systematic trends obtained from CvD and A3D for this common sample of galaxies and discuss possible resolutions. 
Brief conclusions are drawn in Section~\ref{sec:concs}. 

\begin{figure*}
\includegraphics[angle=0,width=175mm]{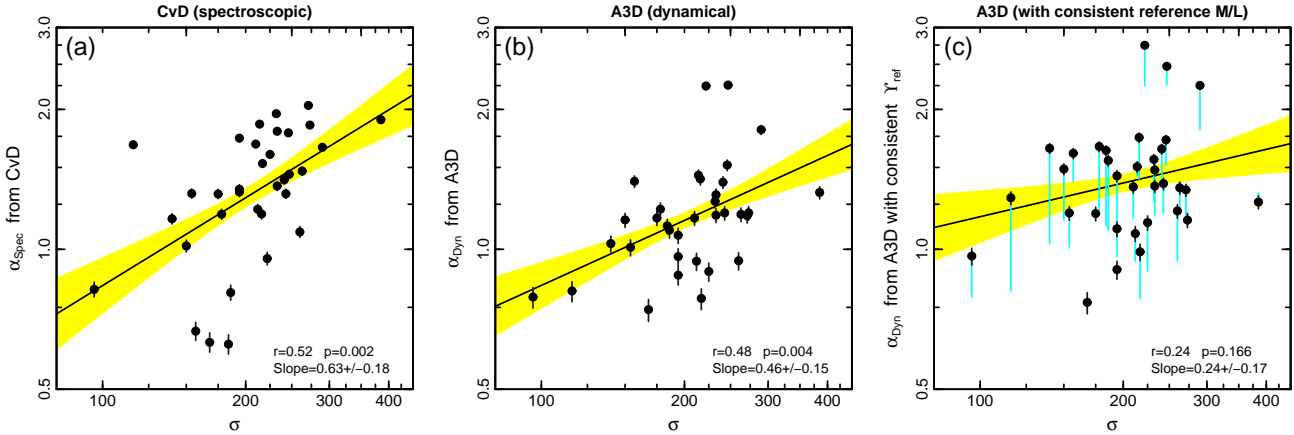}
\vskip -3mm
\caption{The relationship between IMF mismatch factor $\alpha$ and velocity dispersion $\sigma$, as derived from spectroscopy by CvD and from dynamics by A3D,
for the same sample of galaxies. Panel ``c''  shows A3D adjusted to use the CvD-derived reference mass-to-light ratio, for consistency. Slopes and correlation
coefficients are for $\log\alpha$ versus $\log\sigma$.
The solid line and yellow shading show the linear fit with errors to the data in each panel.}
\label{fig:alphasig}
\end{figure*}

\begin{figure*}
\includegraphics[angle=0,width=175mm]{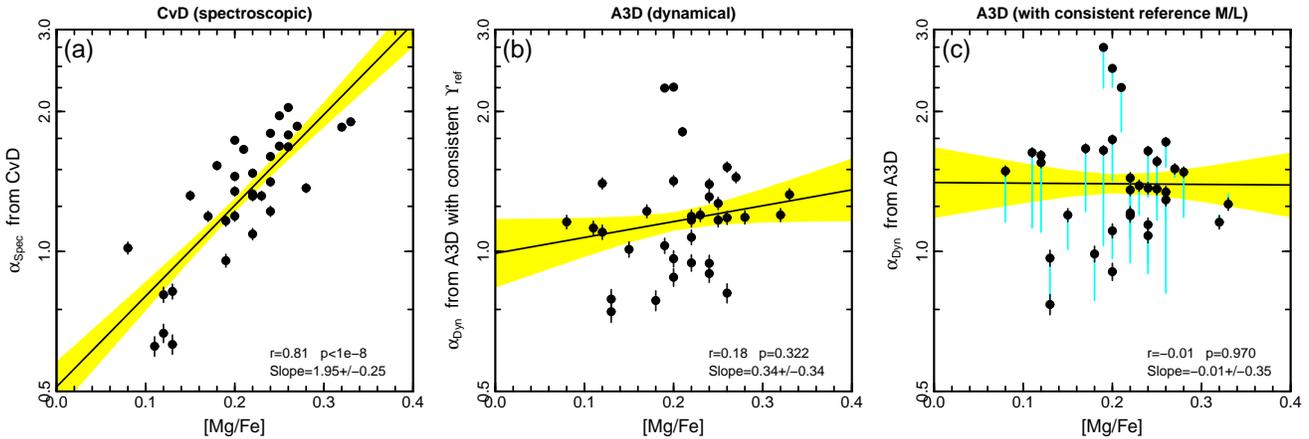}
\vskip -3mm
\caption{Equivalent to Figure~\ref{fig:alphasig} but now for correlations of IMF mismatch factor $\alpha$ with abundance ratio [Mg/Fe].
Slopes and correlation coefficients are for $\log\alpha$ versus [Mg/Fe].
Note how, compared to the previous figure, the correlation is strengthened for CvD but weakened for A3D.
}
\label{fig:alphamgfe}
\end{figure*}

\section{Mass-to-light ratios and IMF factors}\label{sec:mlcomp}

The data employed in this paper are all taken from the published sources (table 2 of CvD and table 1 of A3D).
The key parameters are the stellar mass-to-light ratios $\Upsilon$ derived from spectral fitting by CvD and those inferred by A3D from dynamical modelling, and
the ``reference'' mass-to-light ratios, $\Upsilon_{\rm ref}$, derived by fitting the spectra with models of fixed IMF. 
For clarity, the $\Upsilon_{\rm ref}$ from A3D is converted to a Kroupa (2001) reference IMF (rather than Salpeter 1955), to match the convention 
used by CvD, multiplying by 1.55. The A3D $\Upsilon$ and $\Upsilon_{\rm ref}$ values refer to an aperture of $\sim$1$R_{\rm eff}$ (the effective radius), 
while the CvD quantities are for an aperture of $R_{\rm eff}/8$. (The typical metallicity gradient implies a $\sim$0.2\,dex difference in Z/H between these
apertures, causing only $\sim$10\,per cent difference in mass-to-light ratio.)
Additionally, I use the velocity dispersion and Mg/Fe ratios from CvD to test how the IMF correlates with other key galaxy parameters. 
These quantities are as measured within the central aperture of $R_{\rm eff}/8$. 
From the tabulated parameters, the IMF mismatch factor $\alpha$ is derived by dividing the spectroscopic and dynamical mass-to-light ratios by 
the corresponding reference values, i.e. $\alpha_{\rm Spec}$\,=\,$\Upsilon_{K,{\rm Spec}}^{\rm CvD}/ \Upsilon_{K,{\rm ref}}^{\rm CvD}$ and 
$\alpha_{\rm Dyn}$\,=\,$\Upsilon_{r,\rm Dyn}^{\rm A3D} / \Upsilon_{r,{\rm ref}}^{\rm A3D}$. The difference in passband ($r$ and $K$ subscripts) in these definitions does not
significantly affect the $\alpha$ factor because the additional mass implied by a bottom-heavy IMF (which does not depend on bandpass)
greatly exceeds the additional light (which does).

I begin by comparing the reference mass-to-light values used in the two studies, in Figure~\ref{fig:mlcomp}a.
A version of this comparison was presented by Cappellari et al. (their figure 10), but they compared $K$-band mass-to-light
from CvD with $r$-band from A3D, applying a constant offset (value unstated) to account for the difference in bandpass and reference IMF adopted.
Here, I use the $r$-band $\Upsilon_{\rm ref}$ from CvD, which can be derived from parameters in their table 2, since $\alpha$ is independent of bandpass:
$\Upsilon_{r,{\rm ref}}$\,=\,$\Upsilon_{r,{\rm Spec}} (\Upsilon_{K,{\rm ref}}/ \Upsilon_{K,{\rm Spec}})$.
Whereas A3D found a slope consistent with unity, Figure ~\ref{fig:mlcomp}a, using the $r$-band quantities consistently, shows a signifcant tilt,
with CvD finding progressively smaller $\Upsilon_{\rm ref}$ for lower-$\Upsilon_{\rm ref}$ (younger or more metal-poor)  galaxies. 
The comparison also indicates a zero point shift: the reference mass-to-light ratios used by CvD are on average 20\,per cent smaller than those from A3D
(the $\sim$10\,per cent offset due to metallicity gradients would act in the opposite direction).
The scatter around a linear trend line is only 12\,per cent in $\Upsilon_{\rm ref}$. The formal errors are quoted as 7\,per cent and 2\,per cent in A3D and CvD respectively.

Figure~\ref{fig:mlcomp}b shows the equivalent comparison for the best derived stellar mass-to-light ratios from the two 
works, i.e. $\Upsilon_{r,\rm Dyn}^{\rm A3D}$ versus $\Upsilon_{r,\rm Spec}^{\rm CvD}$. 
A similar figure was shown in CvD (their figure 11), but using the {\it total} dynamical mass-to-light from Scott et al. (2009), i.e. including dark matter. 
By contrast, Figure~\ref{fig:mlcomp}b employs the more recently published mass-to-light ratios from Cappellari et al. (2013), 
which refer to the {\it stellar} component alone, and hence are directly comparable with CvD. 
The clear correlation seen in  Figure~\ref{fig:mlcomp}b is an impressive achievement, since there are essentially
no common ingredients between the A3D dynamical masses and the CvD spectroscopic estimates. 
The scatter in this comparison is $\sim$50\,per cent, greatly in excess of the formal statistical errors of $\sim$7\,per cent reported by each study,
but supporting the statement by CvD that systematic errors in $\Upsilon_{\rm Spec}$ are no more than 50\,per cent.

Of course, the measured mass-to-light ratios reflect not only the IMF but also depend on the metallicity and star-formation history. 
To isolate the IMF effects, Figure~\ref{fig:mlcomp}c shows the comparison between the mismatch factors $\alpha_{\rm Dyn}$ and $\alpha_{\rm Spec}$.
This test does not support any correlation between the IMF constraints obtained from the two methods, at a galaxy by galaxy level
(correlation coefficient $r$\,=\,0.17, exceeded with probability 0.34 under the null hypothesis of no correlation).
Since the two studies used reference mass-to-light ratios $\Upsilon_{\rm ref}$ which differ systematically, it is also helpful to construct 
a ``hybrid'' quantity $\alpha_{\rm Dyn, hyb}$ by dividing the dynamical mass-to-light from A3D by the reference value obtained by CvD, 
i.e. $\alpha_{\rm Dyn, hyb}$\,=\,$\Upsilon_{r,{\rm Dyn}}^{\rm A3D} / \Upsilon_{r,{\rm ref}}^{\rm CvD}$. This parameter isolates the differences between dynamical and spectroscopic
mass-to-light ratios, keeping a consistent treatment of the reference mass-to-light. 
Figure~\ref{fig:mlcomp}d repeats the $\alpha$ comparison using  $\alpha_{\rm Dyn,hyb}$ in place of  $\alpha_{\rm Dyn}$. This shifts A3D to larger
$\alpha$ on average, because the average CvD $\Upsilon_{\rm ref}$ is smaller, but again there is 
no significant correlation between the IMF parameters derived from the two studies ($r$\,=\,0.02).

\section{Systematic IMF trends}\label{sec:imftrends}

Although the  galaxy-by-galaxy comparison in the previous section found no correlation between dynamical and spectroscopic $\alpha$, 
some statistical similarity may still appear when analysing the sample as a whole. For example it is clear that the average mass-to-light ratios in both cases
are generally larger than the average reference values, i.e. both methods indicate a heavier-than-MW IMF,  on average.  
This section compares the data at an intermediate level of detail, through the correlations of $\alpha$ with velocity dispersion and [Mg/Fe], i.e. do the
two methods yield similar systematic {\it trends}, when pooling information over the whole sample?

Figure~\ref{fig:alphasig} shows the $\alpha$-vs-$\sigma$ relation derived from the spectroscopic and dynamical methods. 
The results from both methods, 
indicate a significant increase in $\alpha$ with increasing velocity dispersion, at the $\sim$3$\sigma$ level. The A3D relation is marginally 
shallower ($\Delta\log\alpha/\Delta\log\sigma$\,=\,0.46$\pm$0.15, compared to 0.63$\pm$0.18 for CvD) and is offset to slightly lower $\alpha$. 
Using $\alpha_{\rm Dyn,hyb}$ (i.e. forcing the same reference mass-to-light ratio in both studies) shifts A3D to 
higher normalization and reduces the derived slope further to 0.24$\pm$0.17: consistent with zero, and marginally inconsistent with the CvD slope. 
(The slope determined by Cappellari et al. for the {\it full} A3D sample of 260 galaxies is 0.26$\pm$0.05.)

Figure~\ref{fig:alphamgfe} shows the equivalent comparison for the $\alpha$-versus-[Mg/Fe] relation. 
As noted by CvD this is a much stronger correlation for their dataset than the $\alpha$\,--\,$\sigma$ relation; the trend is significant at $>$7$\sigma$.
Notably, however, the same is {\it not} true for the A3D measurements in the same sample: the trend of $\alpha_{\rm Dyn}$ with Mg/Fe is consistent 
with zero ($\Delta\log\alpha/\Delta{\rm [Mg/Fe]}$\,=\,0.34$\pm$0.34, which can be compared to 1.95$\pm$0.25 for CvD).  
Using $\alpha_{\rm Dyn,hyb}$ flattens the A3D slope further (to --0.00$\pm$0.35).

\begin{figure}
\vskip -1mm
\includegraphics[angle=0,width=84mm]{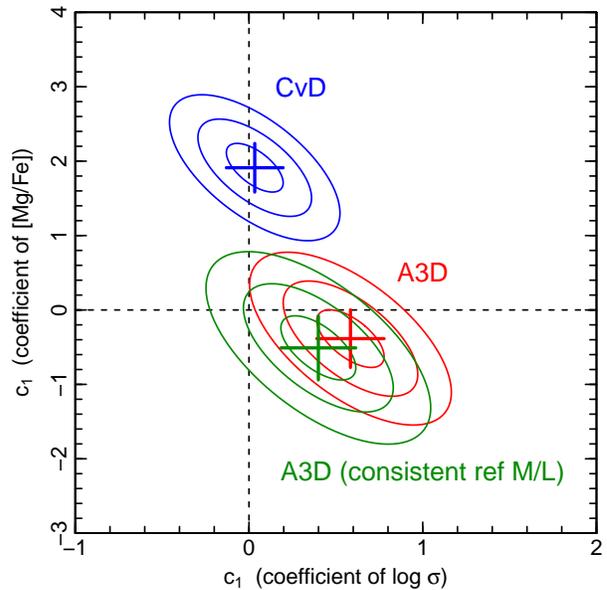}
\vskip -3mm
\caption{Coefficients of bivariate fits to the IMF mismatch factor $\alpha$, as derived for an identical sample of galaxies by CvD (spectroscopic method) 
and ATLAS3D (dynamical). The  contours show the 1,2,3$\sigma$ confidence regions for the coefficients of velocity dispersion and [Mg/Fe], i.e.
$c_1$ and $c_2$ in a model of the form $\alpha = c_0 + c_1 \log(\sigma/200\,{\rm km\,s}^{-1}) + c_2 ([{\rm Mg/Fe}]-0.2)$. 
The dynamical method finds $\alpha$ correlated only with $\sigma$, while the spectroscopic method has $\alpha$ correlated only with abundance ratios.
}
\label{fig:bivcoeff}
\end{figure}

Since Mg/Fe and $\sigma$ are mutually correlated in early-type galaxies, it is helpful to separate the effects of the two parameters using
a simultaneous regression. Fitting a model of the form $\alpha = c_0 + c_1 \log(\sigma/200\,{\rm km\,s}^{-1}) + c_2 ([{\rm Mg/Fe}]-0.2)$, 
I obtain the coefficients and error ellipses which are displayed in Figure~\ref{fig:bivcoeff}. Because  Mg/Fe and $\sigma$  are positively correlated, 
the errors on their coefficients are anti-correlated; nonetheless, there is sufficient scatter around the Mg/Fe\,--\,$\sigma$ relation to identify the dominant statistical 
(though not necessarily causal) ``driver'' of the correlations.  For CvD, the bivariate fit confirms that Mg/Fe is the 
{\it only} informative predictor of $\alpha$: at fixed Mg/Fe, there is {\it no} residual correlation with velocity dispersion. 
By contrast, the A3D measurements show a marginally {\it negative} correlation with Mg/Fe after controlling for the trend with $\sigma$. Despite being based
on the same sample of galaxies, the trends derived from A3D and CvD are different at the $>$4\,$\sigma$ level, based on the error ellipses in Figure~\ref{fig:bivcoeff}.
Using $\alpha_{\rm Dyn,hyb}$ makes little difference to the slopes obtained from the bivariate fit.

\section{Discussion}\label{sec:disc}

Comparing the A3D and CvD determinations of the IMF mismatch factor $\alpha$, for an identical sample of 34 galaxies, I have shown that there is no correlation
between dynamical and spectroscopic $\alpha$, at a galaxy-by-galaxy level, and that the systematic trends with ``mass'' are quite different for the two studies. 
Specifically, although both studies superficially find a correlation of $\alpha$ with velocity dispersion, this trend in CvD merely reflects an underlying correlation with Mg/Fe, 
while for A3D the correlation is only with $\sigma$. 

A sceptical interpretation of these results would be to conclude that possible confounding factors have not been completely accounted for in one or both of the two methods. 
When the IMF is constrained using the strengths of metal lines, it is found to be correlated only with metal abundances. 
When dynamical models are used, the IMF is found to correlate only with a dynamical quantity.
Of course, the IMF {\it could} be intrinsically related either to the velocity dispersion or to the abundance ratios; the problem
is that the two methods, applied to an identical sample, yield incompatible conclusions.
Both CvD and A3D take great care to account for the degeneracies in their methods: 
For the dynamical estimates, Cappellari et al. (2012) explored a wide range of plausible dark-matter halo models, with and without baryonic ``contraction'', and found 
that the requirement for a heavy IMF at high $\sigma$ was robust with respect to the halo prescription.
For spectroscopy, CvD used detailed stellar atmosphere calculations  (Conroy \& van Dokkum 2012a) to disentangle the IMF from effects
of different abundance mixtures on the integrated spectra, and fit the abundance pattern (and other nuisance parameters) simultaneously with the IMF model.
Hence if the discrepancies highlighted in this Letter do arise from incomplete resolution of either
IMF/dark-matter or IMF/abundance degeneracies, the mechanism must be rather subtle.
 
\begin{figure}
\vskip -1mm
\includegraphics[angle=270,width=84mm]{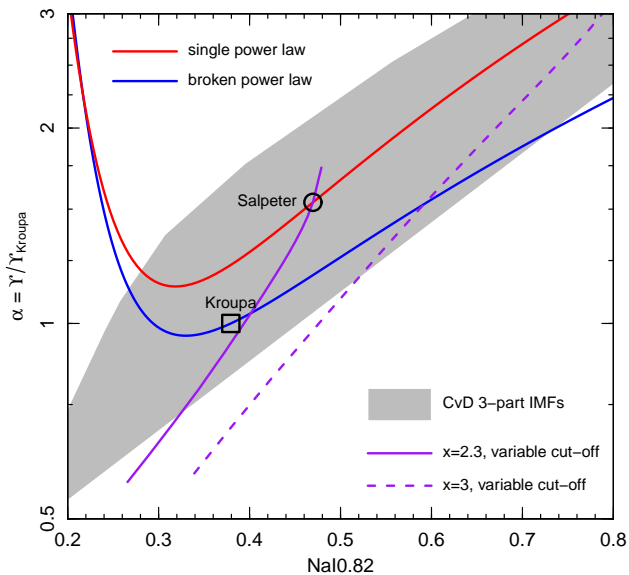}
\vskip -3mm
\caption{
Effect of different assumed IMF models in the conversion from spectroscopic constraints (here represented by the NaI0.82 index defined by Conroy \& van Dokkum 2012a),
to mass-to-light factor $\alpha$\,=\,$\Upsilon/\Upsilon_{\rm ref}$. One-parameter models (e.g. in which a single slope, or a single cut-off mass, is varied) imply a one-to-one 
mapping from spectroscopic information to $\alpha$. Models in which the high-mass slope is varied can become dominated by remnant mass (upper left). 
The three-part IMFs used by CvD have fixed high-mass slope (so are never remnant-dominated), but two free parameters for the low-mass slopes, hence a broader range in $\alpha$ for given spectroscopy.
The purple lines show power-law IMFs with fixed slope but varying low-mass cut-off ($M_{\rm low}$\,=\,0.08--0.5\,$M_\odot$). All other models are integrated over a mass range 0.1--100\,$M_\odot$.
The quantities shown here were computed using a 13.5\,Gyr solar-composition isochrone from the BaSTI database, including low-mass tracks from
Cassisi et al. (2000), and an approximate fitting function for NaI0.82 derived from figure 7 of Conroy \& van Dokkum (2012a).
}
\label{fig:imfmod}
\end{figure}

Alternatively, it is possible that both methods correctly measure different aspects of the IMF. Recall that $\alpha$ refers to a ratio of mass-to-light ratios, but
only the dynamical method strictly measures mass. CvD instead measure the contribution of cool dwarf stars to the integrated luminosity.
The mass-to-light ratio $\Upsilon_{\rm Spec}$ is obtained by fitting a three-part power-law IMF model, and computing the 
mass and luminosity from this model, integrating from 0.1 to 100\,$M_{\odot}$. The result can depend quite sensitively on the form of the shape of the 
adopted model, as shown in Figure~\ref{fig:imfmod}. 
Since the features used by CvD do not constrain the IMF shape in detail,  their $\Upsilon_{\rm Spec}$ are inevitably dependent, to some extent, on the form 
of the model imposed. This point has also been 
emphasised by La Barbera et al. (2013), who note that different prescriptions for IMF models 
(single-slope vs two-part power-laws), which fit galaxy spectra similarly well, can have very different mass-to-light ratios.
Potentially, then, the dynamical measurements of $\alpha$, probing the total stellar mass, are measuring variation in the very-low-mass end of the 
IMF (or even perhaps the contributions of stellar remnants) while spectroscopic $\alpha$ are measuring a different ``moment'' of the IMF. 
Barnab\`e et al. (2013) have used this argument to constrain jointly the slope and cut-off mass for a power-law IMF in lensing ellipticals.
In this scenario, the large scatter observed between spectroscopic and dynamical $\alpha$ would imply a galaxy-to-galaxy variation in the IMF
{\it in excess} of what can be constrained by spectral features, ruling out one-parameter forms for the IMF.
Moreover, the separate parameter dependencies for $\alpha_{\rm Spec}$ and $\alpha_{\rm Dyn}$ could then indicate that different aspects of the IMF correlate with different 
properties, e.g. slope varying with Mg/Fe but low-mass cut-off dependent on $\sigma$. 

A further possibility is that the IMF varies substantially {\it within} galaxies, so that the $R_{\rm eff}/8$ spectroscopic aperture probes populations
with a signficantly different degree of dwarf-enrichment compared to the $\sim$$R_{\rm eff}$ scale probed by the dynamics. 
Spatially-resolved spectroscopic IMF constraints (e.g. Mart\'in-Navarro et al. 2014; Pastorello et al. 2014) should soon be able to test whether this explanation is viable. 

Finally, I note that the strong dependence of $\alpha$ on Mg/Fe found by CvD (and supported by Smith et al. 2012) is apparently {\it not} seen in the 
study of stacked spectra from the Sloan Digital Sky Survey by La Barbera et al. (2013). They found little correlation of IMF-dependent features against Mg/Fe at fixed $\sigma$ 
after correcting for age and metallicity effects (their figure 8). Fitting a broken-power law IMF to composite spectra binned as a function of both $\sigma$ and Mg/Fe, 
they obtain a very strong dependence on $\sigma$ and only a weak correlation with Mg/Fe, the latter simply reflecting the Mg/Fe-vs-$\sigma$ relation (F. La Barbera, private communication).

\section{Conclusions}\label{sec:concs}

I have presented a comparison between spectroscopic (CvD) and dynamical (A3D) results on the IMF in elliptical galaxies, using a common sample
of 34 galaxies with measurements in both studies. 

The analysis shows that ``consensus'' between dynamical and spectroscopic measurements is present only at the most rudimentary level:
both approaches find that a heavier-than-MW IMW is required, on average, for the common sample,
but there is no correlation between the mass-excess factors derived from the two methods, on a galaxy-by-galaxy level.
The two studies apparently find a correlation of $\alpha$ with some quantity related to galaxy mass. 
When plotted only against $\sigma$, there is reasonable agreement in slope, 
but this treatment obscures a clear discrepancy between the results: the correlation found by CvD is not a trend with $\sigma$, but entirely 
with the Mg/Fe abundance ratio. By contrast, A3D finds no correlation of $\alpha$ with Mg/Fe and is hence in significant conflict with the spectroscopic method.

The sense of this disagreement could indicate that confounding factors such as dark matter contributions (A3D) or unusual abundance patterns (CvD)
have not been correctly separated from the IMF effects in one or other of the methods. 
Alternatively, since the two methods are sensitive to different aspects of the IMF, 
further comparison between dynamical and spectroscopic estimates of $\alpha$ 
might lead to a more detailed understanding of the {\it shape} of the IMF and its possible variation in elliptical galaxies. 

Work is ongoing to derive spectroscopic IMF constraints for more galaxies in the A3D sample, and to improve the treatment of 
element abundance treatment in the CvD models (C. Conroy, private communication).  Hence, an enlarged and updated 
comparison between the two methods should be possible in the near future.


\section*{Acknowledgements}

This work was supported by STFC Rolling Grant ST/I001573/1. 
All data used are available in published sources.
I am grateful to John Lucey for helpful discussions and comments on this work, and to 
Francesco La Barbera and the referee Ignacio Ferreras, for updates regarding their 
analysis of SDSS composite spectra.

{}

\label{lastpage}

\end{document}